\documentclass[aps,prd,reprint,onecolumn,notitlepage,nofootinbib]{revtex4-1}

\usepackage{hyperref}
\usepackage{graphicx}
\usepackage{multirow}
\usepackage{amsmath}
\usepackage{bm}
\usepackage{slashed}
\usepackage{epsfig}
\usepackage{amsfonts}
\usepackage{color}
\usepackage{epstopdf}
\usepackage{enumitem}
\usepackage{extarrows}
\usepackage{rotating}
\usepackage{autobreak}
\allowdisplaybreaks
\graphicspath{{figures/}} 

\newcommand{\ben}{\begin{eqnarray}}
\newcommand{\een}{\end{eqnarray}}

\newcommand{\bef}{\begin{figure}[!htp]}
\newcommand{\eef}{\end{figure}}

\newcommand{\bea}{\begin{eqnarray}}
\newcommand{\eea}{\end{eqnarray}}

\def\ba{\begin{linenomath*}\begin{equation}}
\def\ea{\end{equation}\end{linenomath*}}

\allowdisplaybreaks

\newcommand{\state}[4]{{^{#1}\hspace{-0.6mm}#2_{#3}^{[#4]}}}

\newcommand\CSaSz{\state{1}{S}{0}{1}}
\newcommand\CSaPa{\state{1}{P}{1}{1}}
\newcommand\CScSa{\state{3}{S}{1}{1}}

\newcommand\CScPa{\state{3}{P}{1}{1}}
\newcommand\CScPb{\state{3}{P}{2}{1}}
\newcommand\CScPj{\state{3}{P}{J}{1}}
\newcommand\COaSz{\state{1}{S}{0}{8}}
\newcommand\COaPa{\state{1}{P}{1}{8}}
\newcommand\COcSa{\state{3}{S}{1}{8}}
\newcommand\COcPz{\state{3}{P}{0}{8}}

\newcommand\COcPj{\state{3}{P}{J}{8}}

\newcommand\mo{{\mathcal O}}

\newcommand{\vt}[1]{{{\boldsymbol #1}_\perp}}
\newcommand{\vtn}[2]{{{\boldsymbol #1}_{#2\perp}}}

\newcommand{\vp}{{\vt{p}}}
\newcommand{\vk}{{\vt{k}}}
\newcommand{\vka}{{\vtn{k}{1}}}

\newcommand{\vtp}[1]{{{\boldsymbol #1}'_\perp}}

\newcommand{\vkp}{{\vtp{k}}}

\begin{document}
\title{A short theoretical review of charmonium production}

\author{An-Ping Chen$^{a}$}
\email{chenanping@pku.edu.cn}
\author{Yan-Qing Ma$^{b,c,d}$}
\email{yqma@pku.edu.cn}
\author{Hong Zhang$^{e}$}
\email{hong.zhang@sdu.edu.cn}
\affiliation{
        $^{a}$College of Physics and Communication Electronics, Jiangxi Normal University, Nanchang 330022, China\\
       $^{b}$School of Physics and State Key Laboratory of Nuclear Physics and
	   Technology, Peking University, Beijing 100871, China\\
       $^{c}$Center for High Energy physics, Peking University, Beijing 100871, China\\
       $^{d}$Collaborative Innovation Center of Quantum Matter,
	    Beijing 100871, China\\
	   $^{e}$Key Laboratory of Particle Physics and Particle Irradiation (MOE),
Institute of Frontier and Interdisciplinary Science,
Shandong University, (QingDao), Shandong 266237, China
}
\date{\today}

\begin{abstract}
In this paper, we review the current status of the phenomenological study of quarkonium production in high energy collisions. After a brief introduction of several important models and effective field theories for quarkonium production, we discuss the comparisons between theoretical predictions and experimental measurements.
\end{abstract}

\maketitle

\section{Introduction}
\label{Intro}
Since the discovery of the $J/\psi$ in 1974, heavy quarkonium has been on the focus of much experimental and theoretical attention.
Heavy quarkonium is a bound state consisting of a heavy quark ($Q$) and its anti-quark ($\bar Q$).
Depending on the flavor of the quark pair, there are charmonium and bottomonium.
The production of a heavy quarkonium involves three different momentum scales: the heavy quark mass $m_Q$ ($m_c \approx 1.3~\textrm{GeV}$ and $m_b \approx
4.2~\textrm{GeV}$ in $\overline{\textrm{MS}}$ scheme), which governs the perturbative creation of the heavy quark pair ($Q\bar Q$);
the heavy quark momentum $m_Qv$ in the quarkonium rest frame; and the typical
heavy quark kinetic energy $m_Qv^2$, which governs the nonperturbative hadronization
of the $Q\bar Q$ to physical quarkonium. Here $v$ is the typical heavy quark velocity in the quarkonium rest frame ($v^2\approx0.3$ for
charmonium and $v^2\approx0.1$ for bottomonium).
Due to the  non-relativistic nature of the bound state, heavy quarkonium production at high energy collisions is a very important process to test our understanding of QCD.

Experimentally, many quarkonium states are relatively simple to identify in different colliders because of their clean experimental signatures and reasonably high yields.
By virtue of these advantages, the heavy quarkonium is considered as an promising tool to study the inner parton structure of the initial-state hadrons, such as the parton distribution functions (PDFs) and the transverse momentum dependent distributions (TMDs) of proton.
In recent years, heavy quarkonium production is also studied in heavy ion collisions to probe  the quark-gluon plasma (QGP).
The heavy quark pair is first produced in hard scattering at the early stage of the collisions, and then interacts with the QGP and hadronizes to the heavy quarkonium on its way out of the QGP.
Therefore, a good understanding of the heavy quarkonium production mechanism could facilitate our understanding of all these QCD objects.

A lot of data for the quarkonium production in different high energy collisions have been collected. Take $J/\psi$ as an example, the cross section of $e^+e^-\to J/\psi + X$ ($e^+e^-$ collision) has been measured by the Belle and BaBar collaborations, the cross section of $e^+e^-\to e^+e^- + J/\psi + X$ ($\gamma\gamma$ collision) has been measured by DELPHI collaboration at LEP, the yield and polarization of $J/\psi$ production in $ep\to J/\psi + X$ (photoproduction)
have been measured by $\mathrm{H1}$ and Zeus at HERA, and the yield and polarization of $J/\psi$ in hadroproduction ($pp$ or $p\bar p$ collision) have been measured
by CDF at Tevatron, by PHENIX and STAR at RHIC, and by
CMS, ATLAS, ALICE, and LHCb experiments at the LHC.
In the meantime, lots of theoretical efforts have been made to explain these experimental measurements. In this article, we will briefly review the current status of the phenomenological study of quarkonium production. We first overview theoretical frameworks for describing the inclusive quarkonium production. Then we review the current status of the comparison between theoretical predictions and experimental measurements. We put stress on charmonium  production at hadron colliders, with brief overview of the quarkonium production in $e^+e^-$, $ep$ and $\gamma\gamma$ collisions. We refer the readers to several relevant reviews \cite{Chapon:2020heu,Lansberg:2019adr,Andronic:2015wma,Brambilla:2010cs,Chung:2018lyq} for detailed
discussions of other topics in quarkonium physics.

\section{Quarkonium production mechanism}

\label{baseline}
 Heavy quarkonium production is usually separated into two steps: (1) the production of a $Q\bar Q$ pair with definite spin and color state in a hard collision, which could be calculated perturbatively; and (2) hadronization of the $Q\bar Q$ pair into a physical heavy quarkonium at a momentum scale much less than the heavy quark mass $m_Q$, which is in principle nonperturbative. Different treatments of the nonperturbative transition from $Q\bar Q$ pair to the physical quarkonium lead to different theoretical models. In the
following, we briefly describe some of the most widely-used ones: the color evaporation model (CEM)~\cite{Fritzsch:1977ay,Gluck:1977zm,Barger:1979js}, the color singlet model (CSM)~\cite{Ellis:1976fj,Carlson:1976cd,Chang:1979nn}, the non-relativistic QCD (NRQCD) factorization theory~\cite{Bodwin:1994jh}, the fragmentation function approach~\cite{Kang:2014tta,Kang:2014pya,Kang:2011mg,Kang:2011zza,Fleming:2012wy}, and the most recently proposed soft gluon factorization (SGF) approach~\cite{Ma:2017xno,Chen:2020yeg}.

\paragraph{The color evaporation model (CEM)}

In the CEM~\cite{Fritzsch:1977ay,Gluck:1977zm,Barger:1979js}, it is assumed that every
produced $Q\bar Q$ pair evolves into a specific heavy quarkonium if its
invariant mass is below the open-charm/bottom threshold. It is further
assumed that the probability for the $Q\bar Q$ pair to evolve into a specific quarkonium state $H$ is given by a
constant $F_H$ which is independent of momentum and process. Mathematically, the production cross section of $H$ is expressed in the CEM as
\begin{align}\label{eq:CEM}
\sigma_H = F_H\int_{2m_Q}^{2m_D} \frac{d \sigma_{Q\bar Q}}{dm_{Q\bar Q}} dm_{Q\bar Q},
\end{align}
where $2m_D$ is the open-charm/bottom threshold. For each heavy quarkonium state, the CEM in Eq.~\eqref{eq:CEM} has one free parameter $F_H$.
The CEM is intuitive, simple, and successful to
explain $J/\psi$ production data. However, it has a very strong prediction that the production rate of any two different charmonium states depends on neither the process nor kinematic variables, which contradicts data from many experiments.
For example, the ratio of production cross section of $\psi(2S)$ to that of $J/\psi$ in $pp$ collisions clearly depends on their transverse momentum~\cite{Adare:2011vq,Aaij:2012ag}.
To overcome these obstacles, an improved version of the model, the ICEM, was proposed~\cite{Ma:2016exq}, in which the momentum of ${Q\bar Q}$ pair is assumed to be larger than the momentum of quarkonium $H$ by a factor of $m_{Q\bar Q}/m_H$. One consequence is that  the lower limit of the above integral is replaced by $m_H$.
It was shown that the ICEM can describe the charmonium
yields as well as the ratio of $\psi(2S)$ over $J/\psi$~\cite{Ma:2016exq}. The ICEM was also combined with $k_T$-factorization to describe quarkonium polarization \cite{Cheung:2017loo,Cheung:2017osx,Cheung:2018tvq,Maciula:2018bex,Cheung:2018upe,Cheung:2021epq}.

\paragraph{The color singlet model (CSM)}

In the CSM, the $Q\bar Q$ pair that evolves into the quarkonium is assumed
to have the same color, spin and orbital-angular-momentum quantum numbers as the heavy quarkonium. Particularly, it must be in a color singlet state. Under this assumption, the production cross section for each quarkonium state $H$ is related to the wave-function (or its derivatives) of $H$ around the origin, which can be extracted from the decay process of $H$, or calculated from the potential model or lattice QCD. Therefore, the CSM effectively has no free parameters. At relatively low energies,
the LO CSM predictions for quarkonium production agree with the experimental data. While at high energies, the LO CSM predictions have been shown to underestimate the experimental data of direct $J/\psi$ and $\psi(2S)$ production at $\sqrt{s}=1.8$~TeV $pp$ collisions \cite{Abe:1997jz} by more than an order of magnitude, which is known as {\it the $\psi(2S)$ surplus puzzle}. In the past decade, it was found that the NLO and NNLO corrections to the CSM are significantly larger than the
LO contributions~\cite{Artoisenet:2007xi,Campbell:2007ws,Artoisenet:2008fc}. Including these corrections relieves the inconsistency between the LO CSM prediction and the data. However, a full description of
data is still difficult. Besides, given the very large corrections at NLO and NNLO, it is not clear that the perturbative expansion in $\alpha_s$ is convergent. Moreover, in the case of $P$-wave production and decay, the CSM is known to be incomplete because it suffers from uncanceled infrared divergences.
The last point can be rigorously cured in a more general
framework of NRQCD factorization theory which we will discuss below.

\paragraph{The NRQCD factorization approach}

NRQCD
is an effective theory of QCD and reproduces full QCD
dynamics at momentum scales of order $m_Qv$ and smaller. In NRQCD, the production cross section of a heavy quarkonium $H$ is given by the
factorization formula~\cite{Bodwin:1994jh}
\begin{align}\label{eq:NRQCD}
\sigma_H = \sum_n  \sigma_{n}(\mu_\Lambda) \langle \mo^{H}_n(\mu_\Lambda)\rangle.
\end{align}
Here $\mu_\Lambda$ is the NRQCD factorization scale, which is the ultraviolet (UV) cutoff of the NRQCD effective theory, $\sigma_n$ is the short-distance coefficient (SDC) which describes the production of a $Q\bar Q$ pair with quantum number $n$ in the hard scattering, and $\langle \mo^{H}_n(\mu_\Lambda)\rangle$ is the NRQCD long-distance matrix element (LDME) that describes the hadronization of the $Q\bar Q$ pair in state $n$ into the heavy quarkonium $H$. The LDME is defined as the vacuum expectation value of a four-fermion operator in NRQCD, and each LDME has a known scaling behavior in powers of $v$.
Then the sum over $n$ can be organized in powers of $v$. Therefore Eq.~\eqref{eq:NRQCD} is a double expansion of $\alpha_s$ and $v$. In practice, for a certain accuracy, one truncates the summation and keeps only a few LDMEs for each $H$ production.
The predictive power of the NRQCD factorization approach relies on the  convergence of this velocity expansion, as well as the universality of LDMEs.

As shown in Eq.~\eqref{eq:NRQCD}, the NRQCD factorization contains contributions from both the color-singlet (CS) and the color-octet (CO) channels.
If one sets the CO contributions to zero, one could recover the CSM for S-wave heavy quarkonium production.
Thanks to
the CO contributions, NRQCD solves the infrared divergence problem encountered in the CSM~\cite{Bodwin:1992qr}.
Although there is no all-order proof of
NRQCD factorization for quarkonium production yet,
it is
found that the factorization holds at least to next-to-next-to-leading
order (NNLO) in $\alpha_s$ if the LDMEs are modified to be gauge complete \cite{Nayak:2005rt,Nayak:2005rw,Nayak:2006fm,Bodwin:2019bpf,Zhang:2020atv}.

\paragraph{The fragmentation function approach}

The SDC's in the NRQCD factorization formula in Eq.~\eqref{eq:NRQCD} suffer from large high-order $\alpha_s$ corrections for heavy quarkonium
produced at large transverse momentum $p_T\gg m_Q$,
which is an important kinematic region in high-energy colliders.
In this region, the high-order corrections of the SDC's receive huge power enhancement in terms of $p_T^2/m_Q^2$, as well as large logarithmic corrections in terms of $\textrm{ln}(p_T^2/m_Q^2)$.
To overcome these problems, a new QCD factorization approach which combined the fragmentation function (FF) approach and NRQCD factorization approach (FF+NRQCD) was proposed to describe the large $p_T$ heavy quarkonium production~\cite{Kang:2014tta,Kang:2014pya,Kang:2011mg,Kang:2011zza,Fleming:2012wy}.
In the FF+NRQCD factorization approach, the cross section is first expanded by powers of $m_Q^2/p_T^2$.
Both the leading-power (LP) term and next-to-leading-power (NLP) term of the expansion could be factorized systematically into parton-production cross sections convoluted with several universal FFs, i.e.
\begin{align}\label{eq:FFNRQCD}
d\sigma_{A+B\to H+X}(p_T) = &
 \sum_{f}
d\hat{\sigma}_{A+B\to f+X}(p_T/z,\mu_F)
\otimes D_{f\to H}(z,m_Q,\mu_F) \nonumber\\
 &\hspace{-2cm}+  \sum_{\kappa}
d{\hat{\sigma}}_{A+B\to [Q\bar{Q}(\kappa)]+X}(P_{[Q\bar{Q}(\kappa)]}=p_T/z,\mu_F) \otimes {\cal
D}_{[Q\bar{Q}(\kappa)]\to H}(z,m_Q,\mu_F)
\nonumber\\
 &\hspace{-2cm} +\mo(m_Q^4/p_T^4)\,,
\end{align}
where the first term on the right side gives the contribution
of LP in $m_Q^2/p_T^2$, and the second term gives the
NLP contribution.
The symbol $\otimes$ represents the convolution of the light-cone momentum fraction $z$.
In the first term,
$d{\hat{\sigma}}_{A+B\to f+X}$ is the semi-inclusive cross section for initial hadrons $A$ and $B$ to produce an on-shell parton $f$ .
The FFs $D_{f\to H}$ represents the possibility of finding $H$ in the hadronization products of  parton $f$.
The NLP contribution is similar, with an intermediate heavy quark pair in state $\kappa$ instead of a single parton $f$.

Since the NLP term is suppressed by $m_Q^2/p_T^2$, it seems to be not important at large $p_T$.
However, it is natural to expect a heavy quark pair is more likely to evolve into a heavy quarkonium $H$, comparing to a single quark or gluon.
Consequently, the double-parton FFs are more important than the single-parton ones.
This nonpertubative enhancement could balance the perturbative $m_Q^2/p_T^2$ suppression in the intermediate $p_T$ range.
There are more NLP contributions from other intermediate double partons besides a heavy quark pair.
They are not included in Eq.~\eqref{eq:FFNRQCD} because they do not have this nonperturbative enhancement.

The factorization formula in Eq.~\eqref{eq:NRQCD} is a double expansion of $m_Q^2/p_T^2$ and $\alpha_s$.
The factorization scale $\mu_F$ is chosen at the same order of $p_T$ so no large logarithm exists.
The FFs with different $\mu_F$ are related by a closed set of evolution equations.
To make a theoretical prediction, one still needs a set of input FFs at a certain scale $\mu_0\sim 2m_Q$.
These input FFs are nonperturbative and, in principle, should be extracted from fitting experimental data.
However, since $\mu_0\sim 2m_Q\gg \Lambda_{\textrm{QCD}}$, it is natural to use NRQCD factorization to further factorize these input FFs.
By doing this, all unknown input FFs could be expressed in terms of NRQCD LDMEs with the perturbatively-calculable coefficients
\begin{subequations}\label{eq:NRQCD-form}
\begin{align}
D_{f\to H}(z,m_Q,\mu_0) =& \sum_n \hat{d}_{f \to Q\bar{Q}[n]}(z,m_Q,\mu_0,\mu_\Lambda) \langle \mathcal{O}_{n}^{H}(\mu_\Lambda)\rangle\, ,\\
{\cal D}_{[Q\bar{Q}(\kappa)] \to H}(z,m_Q,\mu_0)=&\sum_n \hat{d}_{[Q\bar{Q}(\kappa)] \to Q\bar{Q}[n]}(z,m_Q,\mu_0,\mu_\Lambda) \langle \mathcal{O}_{n}^{H}(\mu_\Lambda)\rangle\, .
\end{align}
\end{subequations}
where $\hat{d}_{f \to Q\bar{Q}[n]}$ and
$\hat{d}_{[Q\bar{Q}(\kappa)] \to Q\bar{Q}[n]}$ describe the perturbative
evolution of a parton $f$ and a  $Q\bar Q$ pair with quantum number $\kappa$ into a
$Q\bar Q$ pair in the state $n$, respectively.
Mathematically, the FF+NRQCD factorization formula in Eqs. \eqref{eq:FFNRQCD} and \eqref{eq:NRQCD-form} is a
reorganization of terms in Eq.~\eqref{eq:NRQCD}.
Physically, the FF+NRQCD factorization method correctly includes the evolution of a heavy quark pair when the relative velocity in the quarkonium rest frame is not much smaller than 1 and the NRQCD does not apply.

During the past two decades, the FFs in Eqs.~\eqref{eq:NRQCD-form} have been widely studied.
The coefficients for all double parton FFs to both $S$-wave and $P$-wave states are calculated up to $O(\alpha_s)$  in refs. \cite{Ma:2013yla,Ma:2014eja,Ma:2015yka}.
The coefficients for all single parton FFs are available up to $O(\alpha_s^2)$\cite{Beneke:1995yb,Braaten:1993mp,Braaten:1993rw,Cho:1994gb,Braaten:1994kd,Ma:1995vi,Braaten:1996rp,Braaten:2000pc,Hao:2009fa,Jia:2012qx,Bodwin:2014bia}
(see \cite{Ma:2013yla,Ma:2014eja,Ma:2015yka} for a summary and comparison).
At $\alpha_s^3$ order, the coefficients of gluon FFs  to $Q\bar{Q}(\CScSa)$, $Q\bar{Q}(\CSaPa)$ , $Q\bar{Q}(\state{1}{S}{0}{1,8})$ and $Q\bar{Q}(\state{3}{P}{J}{1,8})$ are calculated  in refs. \cite{Zhang:2017xoj,Braaten:1993rw,Braaten:1995cj,Bodwin:2003wh,Bodwin:2012xc, Sun:2018yam, Zhang:2018mlo,Artoisenet:2018dbs,Feng:2018ulg, Zhang:2020atv}.
Recently, the heavy quark FFs to $\state{1}{S}{0}{1,8}$ state are obtained in refs. \cite{Zheng:2021ylc,Feng:2021uct}.

With the input FFs and the evolution equations, FF+NRQCD factorization formalism provides a systematic reorganization of the cross section in term of powers of $m_Q^2/p_T^2$ and a systematic method for resumming the potentially large $\textrm{ln}(p_T^2/m_Q^2)$-type logarithms. It is expected to have a better convergence in the $\alpha_s$ expansion than NRQCD.

\paragraph{The soft gluon factorization approach}

As we will discuss later, the NRQCD factorization still encounters some difficulties in describing inclusive quarkonium production data. It is known long time ago that NRQCD have bad convergence in velocity expansion~\cite{Mangano:1996kg}, which may be responsible to the phenomenological difficulties. The aim of SGF is to provide a framework with better convergence~\cite{Ma:2017xno}.

In SGF approach, the differential cross section of the quarkonium $H$ production is factorized as
\begin{align}\label{eq:fac4d}
 (2\pi)^3 2 P_H^0 \frac{d\sigma_H}{d^3P_H}\approx \sum_{n} \int \frac{d^4 P}{(2\pi)^4} {\cal H}_{n}(P) F_{n\to H}(P,P_H),
\end{align}
where $P$ is the momentum of the intermediate $Q\bar Q$ pair, $P_H$ is the momentum of $H$, $F_{n\to H}(P,P_H)$ is soft gluon distribution function (SGD), which describes the hadronization of the $Q\bar Q$ pair into physical quarkonium $H$ by emitting soft hadrons. To account for the effect of soft hadrons emission, which
are mainly soft gluons perturbatively, the momentum of the intermediate state $P$ is kept different from the observed quarkonium momentum $P_H$, which is different
from the treatment in NRQCD. The SGD is defined by QCD fields in small loop momentum region. With an explicit definition of the small region and taking advantage of equations of motion, the SGF is shown to be equivalent to the NRQCD factorization~\cite{Chen:2020yeg}. Nevertheless, comparing with NRQCD, the SGF resums a series of relativistic corrections originating from kinematic effects, which results in a better convergence in velocity expansion. It is expected that the SGF approach may provide a better description of experimental data.

 The first phenomenological application of the SGF approach was carried out in ref. \cite{Li:2019ncs} for exclusive quarkonium production processes. It was shown there that, for $\eta_c+ \gamma$ production at B factories, the SGF provides the best description of experimental data among all existed theoretical calculations.
Recently, quarkonium fragmentation function in SGF has been calculated to NLO in ref. \cite{Chen:2021hzo}, which is the first step to apply SGF to inclusive quarkonium production processes.  With explicit NLO calculation, it was demonstrated that the SGF is valid at NLO level. Phenomenological application for inclusive quarkonium production in the SGF approach is still missing.

\section{Quarkonium production in $pp$ collisions}

\subsection{High $p_T$ heavy quarkonium production}

Based on the NRQCD factorization framework, the heavy quarkonium production in $pp$ collisions has been widely studied. In the large $p_T$ region, the differential cross section of the quarkonium $H$ production can be factorized as
\begin{align}\label{eq:NRQCD-pp}
d\sigma_{A+B\rightarrow H+X} &=\sum_n  d\hat\sigma[n] \langle \mo^{H}_n\rangle \nonumber\\
 &=\sum_{i,j,n}\int dx_1dx_2  f_{i/A}(x_1,\mu_F)f_{j/B}(x_2,\mu_F) d\hat\sigma_{i+j\rightarrow Q\bar Q[n]+X}(\mu_F,\mu_\Lambda) \langle \mo^{H}_n(\mu_\Lambda)\rangle,
\end{align}
where $f$'s are the parton distribution functions (PDFs) for the partons in the initial colliding protons, $\mu_F$ is the collinear factorization scale, and $d\sigma_{i+j\rightarrow Q\bar Q[n]+X}$ is the partonic differential cross section.

At LO in $\alpha_s$, the partonic differential cross sections of  $n=\COcSa,\COaSz,\COcPj$ are scaled as $p_T^{-4},p_T^{-6}$ and $p_T^{-6}$, respectively, which are more important than the CS contribution $n=\CScSa$, scaled as $p_T^{-8}$. By including the CO contributions, the $\psi(nS)$ surplus puzzle in CSM is solved naturally \cite{Kramer:2001hh}. But NRQCD factorization encounters difficulties with charmonium polarizations.
Since the dominant contribution is from the transverse  $\COcSa$ channel,
LO NRQCD predicts that $\psi(nS)$ and $\Upsilon(nS)$ produced at hadron colliders are mainly transversely polarized \cite{Cho:1994ih,Beneke:1995yb,Braaten:1999qk}.
On the contrary, experimental measurements at Tevatron and LHC find these states are almost produced unpolarized \cite{Abulencia:2007us,Abelev:2011md,Aaij:2013nlm,Chatrchyan:2013cla,Chatrchyan:2012woa}.
In addition, the LO calculation in NRQCD is difficult to explain the observed cross section ratio $R_{\chi_{c}}=\sigma_{\chi_{c2}}/\sigma_{\chi_{c1}}$ of the $P$-wave charmonia $\chi_{cJ}$ at Tevatron \cite{CDF:2007mqb}. As the $\COcSa$ channel dominance predicts the ratio $R_{\chi_{c}}$ to be $5/4$ by spin counting \cite{Cho:1995vh,Kniehl:2003pc}, which is much larger than the measured value of $0.75$.

In the past decade, the perturbative SDCs for quarkonium production cross sections and polarizations have been calculated to NLO by three groups \cite{Ma:2010yw,Ma:2010jj,Chao:2012iv,Butenschoen:2010rq,Butenschoen:2012px,Gong:2012ug,Gong:2013qka}.
At this order, the $\COaSz$ and $\COcPj$ channels are $p_T^{-4}$ scaling, so their contributions are also important at large $p_T$.
Even though the NLO SDC's obtained by the three groups agree, they give very different predictions  for $J/\psi$ polarization, due to the different methods used in fitting the CO LDMEs,.

\begin{figure}[htb!]
 \begin{center}
 \vspace*{0.8cm}
 \hspace*{-5mm}
 \includegraphics[width=0.45\textwidth]{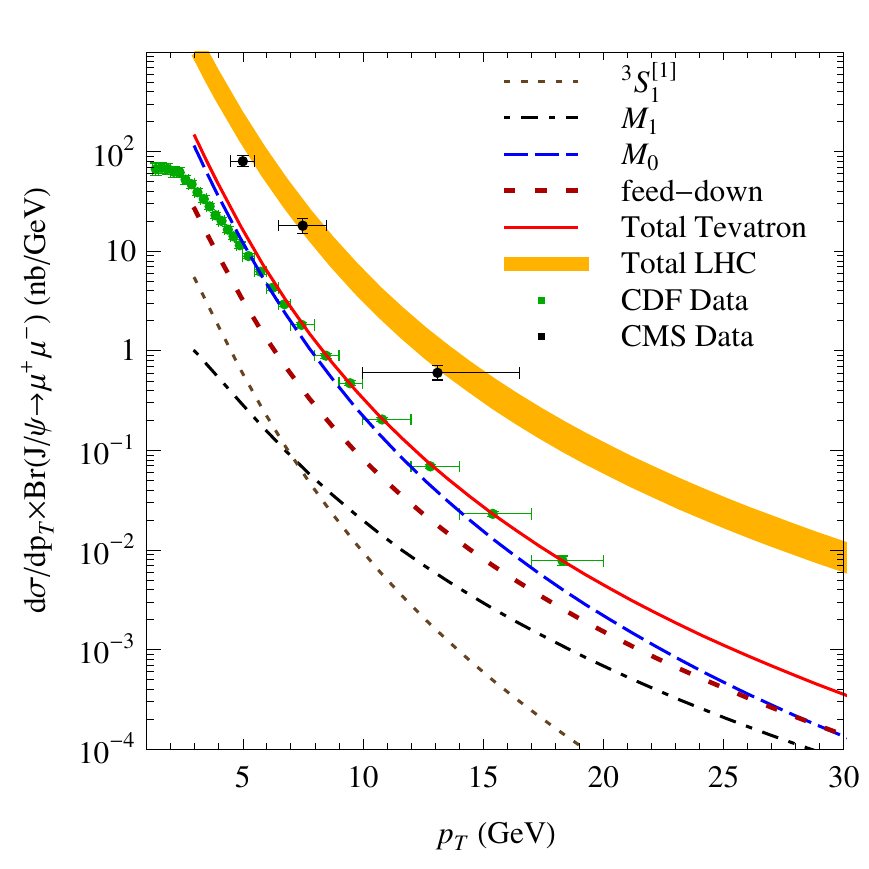}
 \hspace*{5mm}
 \includegraphics[width=0.45\textwidth]{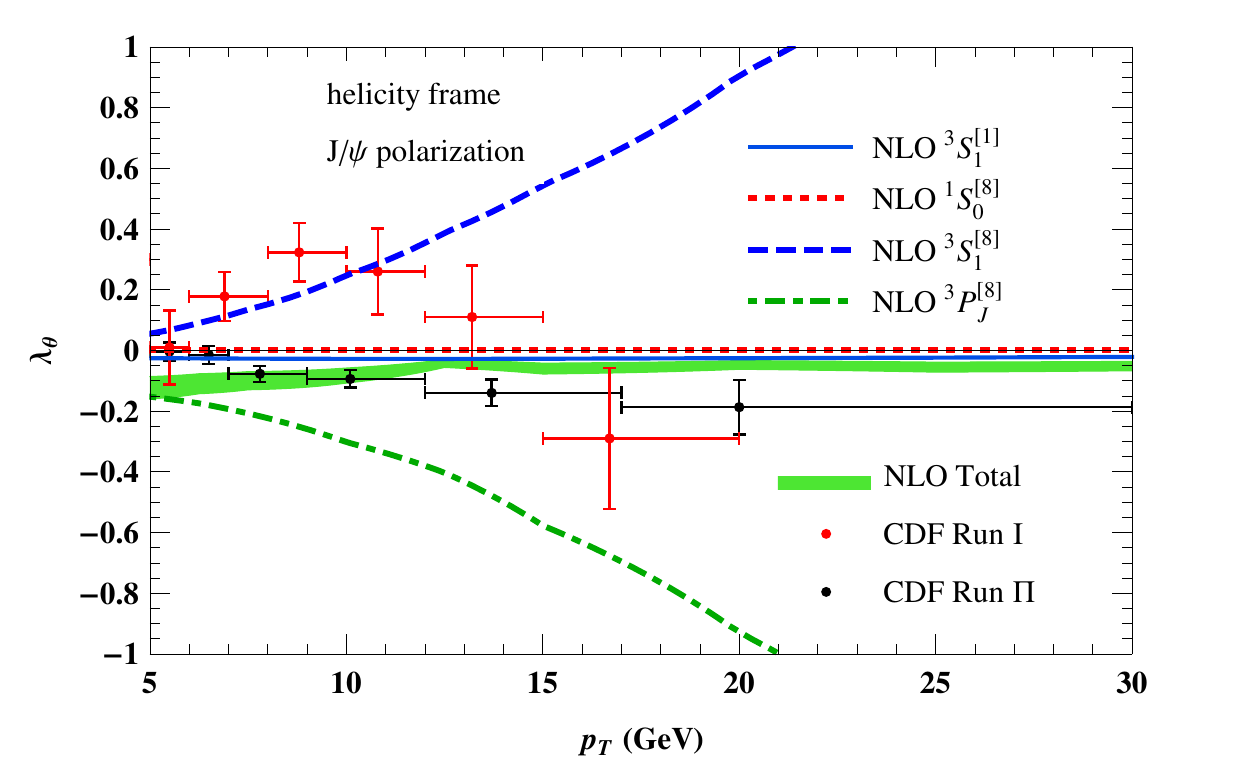}
 \end{center}
 \vspace*{-.5cm}
 \caption{Comparison of NLO NRQCD calculations with prompt $J/\psi$ data. Left panel: The $p_T$ distributions of prompt $J/\psi$ production at the Tevatron and the LHC.   Right panel: NLO NRQCD prediction for the polarization of $J/\psi$ production at the Tevatron. Figures taken from
Ref. \cite{Ma:2010yw,Chao:2012iv}. \label{fig:jpsi}}
 \vspace*{0.cm}
\end{figure}

In refs. \cite{Ma:2010yw,Ma:2010jj,Chao:2012iv}, it was found that at high $p_T$ the SDC of $\COcPj$ channel can be decomposed into a linear combination of the other two CO channels
\begin{align}\label{eq:decompose}
d\hat\sigma[\COcPj] &=r_0 d\hat\sigma[\COaSz] + r_1 d\hat\sigma[\COcSa] ,
\end{align}
where $r_0 = 3.9, r_1 = -0.56$ for the Tevatron, and $r_0 = 4.1, r_1 = -0.56$ for the
LHC. Therefore, two linearly combined NRQCD LDMEs are introduced to fit the data,
\begin{align}\label{eq:combined-LDME}
M_{0,r_0}^H &=\langle \mo^{H}(\COaSz)\rangle + \frac{r_0}{m_c^2} \langle \mo^{H}(\COcPz)\rangle , \\
M_{1,r_1}^H &=\langle \mo^{H}(\COcSa)\rangle + \frac{r_1}{m_c^2} \langle \mo^{H}(\COcPz)\rangle,
\end{align}
where $H$ is $J/\psi$ or $\psi(2S)$. Using the Tevatron data \cite{CDF:2000pfk,Pelaez:2015qba} with $p_T>7$~GeV, the LDMEs are extracted as
\begin{align}
M_{0,r_0}^{J/\psi} &=(7.4\pm 1.9)\times 10^{-2} \mathrm{Gev}^3 , \\
M_{1,r_1}^{J/\psi} &=(0.05\pm 0.02)\times 10^{-2} \mathrm{Gev}^3,
\end{align}
With these fitted LDMEs, the theorical predictions for prompt $J/\psi$ production are generally consistent with experimental data from the Tevatron and the LHC \cite{CDF:2000pfk,Pelaez:2015qba,Hu:2017pat} (see
Fig. \ref{fig:jpsi}).
As shown in the right panel, the $J/\psi$ polarization is roughly
consistent with the CDF data, in which the  $J/\psi$ is approximately
unpolarized.
Detailed analysis shows that the transversely polarized contributions from
$\COcSa$ and $\COcPj$ channels cancel each other.
Thus a possible mechanism is that the unpolarized $\COaSz$ channel dominates, which leads to an unpolarized production \cite{Chao:2012iv, Bodwin:2014gia,Faccioli:2014cqa}.
For the $\psi(2S)$ production, such cancellation is weak \cite{Shao:2014yta}, and its polarization is still hard to explain.

\begin{figure}[htb!]
 \begin{center}
 \vspace*{0.8cm}
 \hspace*{-5mm}
 \includegraphics[width=0.45\textwidth]{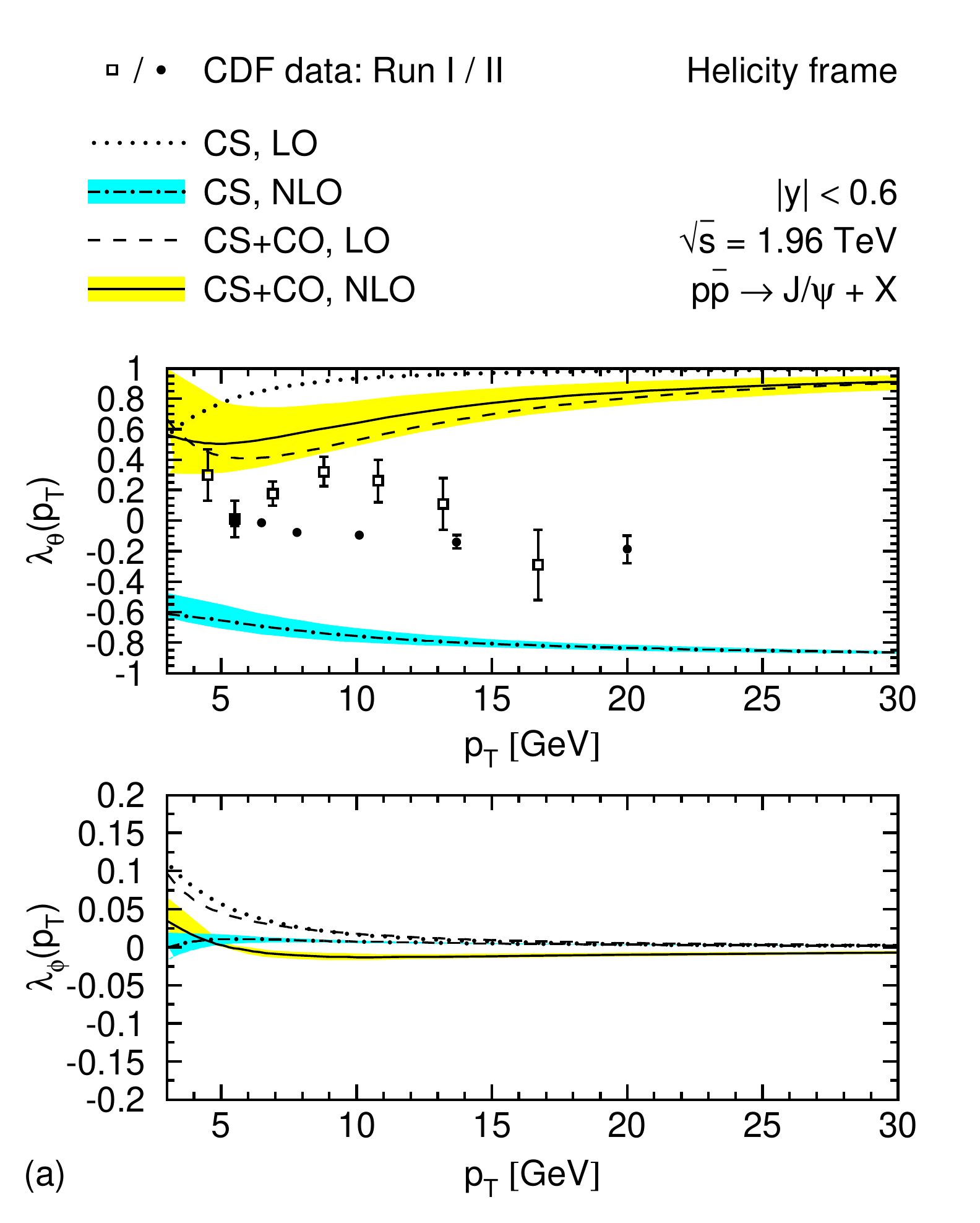}
 \end{center}
 \vspace*{-.5cm}
 \caption{Comparisons of the NLO NRQCD calculations in ref. \cite{Butenschoen:2012px} for the $J/\psi$ polarization with the Tevatron data. Figure from
ref. \cite{Butenschoen:2012px}. \label{fig:jpsi-Kniehl}}
 \vspace*{0.cm}
\end{figure}

In ref. \cite{Butenschoen:2012px}, the authors determine the CO LDMEs by a global fit of a number of measurements. Using prompt $J/\psi$ production data in $pp$ ($p_T>3~\mathrm{GeV}$) \cite{CDF:1997ykw,CDF:2004jtw,PHENIX:2009ghc,CMS:2010nis,ATLAS:2011aqv,Scomparin:2011zzb,LHCb:2011zfl}, $ep$ ($p_T>1$~GeV) \cite{ZEUS:2002src,H1:2002voc,H1:2010udv}, $\gamma\gamma$ \cite{DELPHI:2003hen}, and $e^+e^-$ \cite{Belle:2009bxr} collisions, they determine all three CO LDMEs.
Especially, they obtained a negative value for $\langle \mo^{J/\psi}(\COcPz)\rangle$, and thus the contributions from $\COcSa$ and $\COcPj$ channels add and enhance the transverse polarization at large $p_T$.
With these inputs, it was found that the $J/\psi$ is transversely polarized at large $p_T$ as shown in Fig. \ref{fig:jpsi-Kniehl}.

In ref. \cite{Gong:2012ug}, the LDME determination is based on the yield data from CDF \cite{CDF:2004jtw} and LHCb \cite{LHCb:2011zfl}. Similar to the method in Refs. \cite{Ma:2010yw,Chao:2012iv,Shao:2014yta}, the data with $p_T<7~\mathrm{GeV}$ are not considered in the fitting. It is found that both $\langle \mo^{J/\psi}(\COcSa)\rangle$ and $\langle \mo^{J/\psi}(\COcPj)\rangle$ are negative. As the two LDMEs have same sign, the cancellation between transversely polarized contributions still occurs, which explains the unpolarized $J/\psi$ produced at large $p_T$.

\begin{figure}[htb!]
 \begin{center}
 \vspace*{0.8cm}
 \hspace*{-5mm}
 \includegraphics[width=0.3\textwidth]{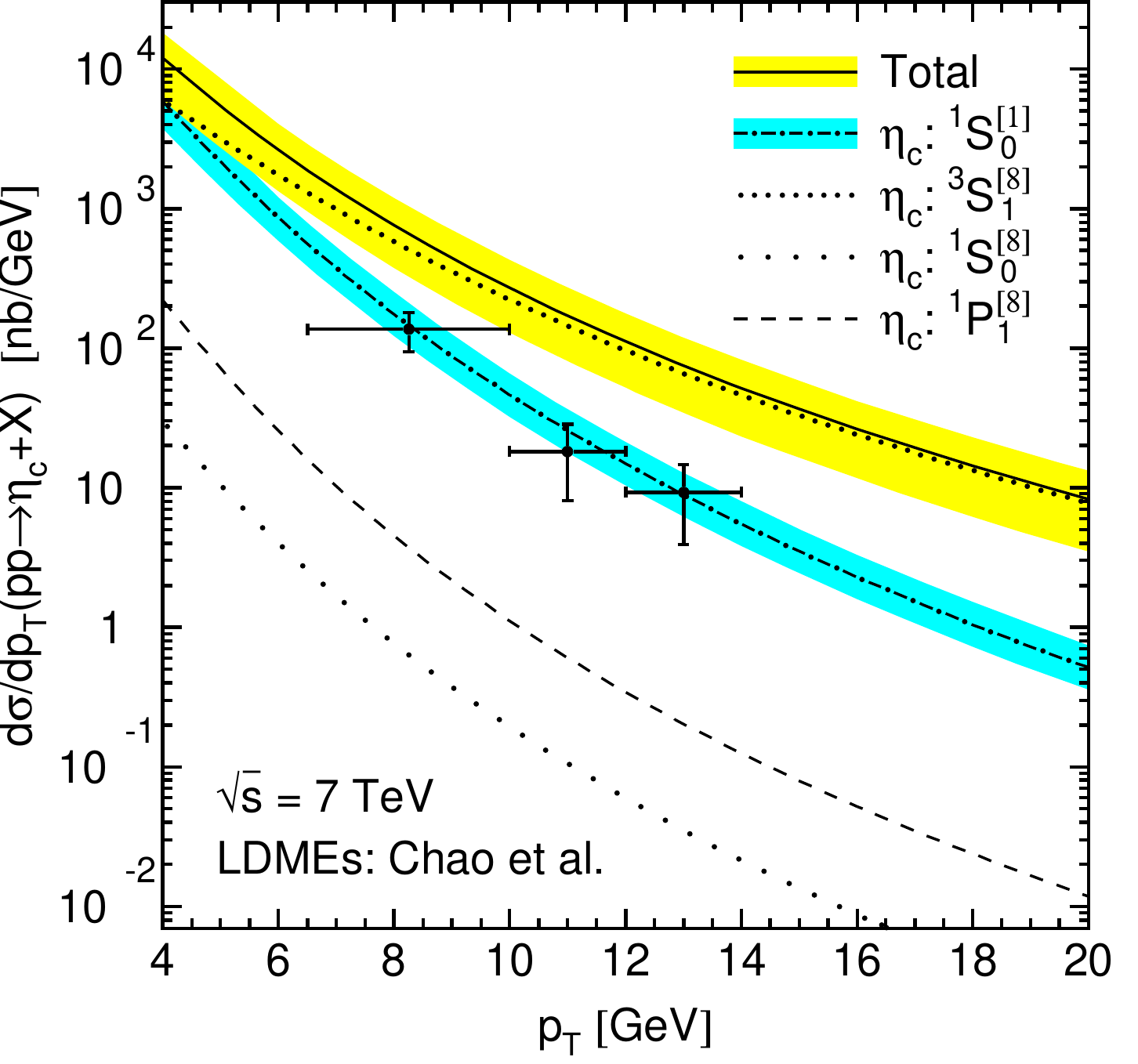}
 \hspace*{5mm}
 \includegraphics[width=0.3\textwidth]{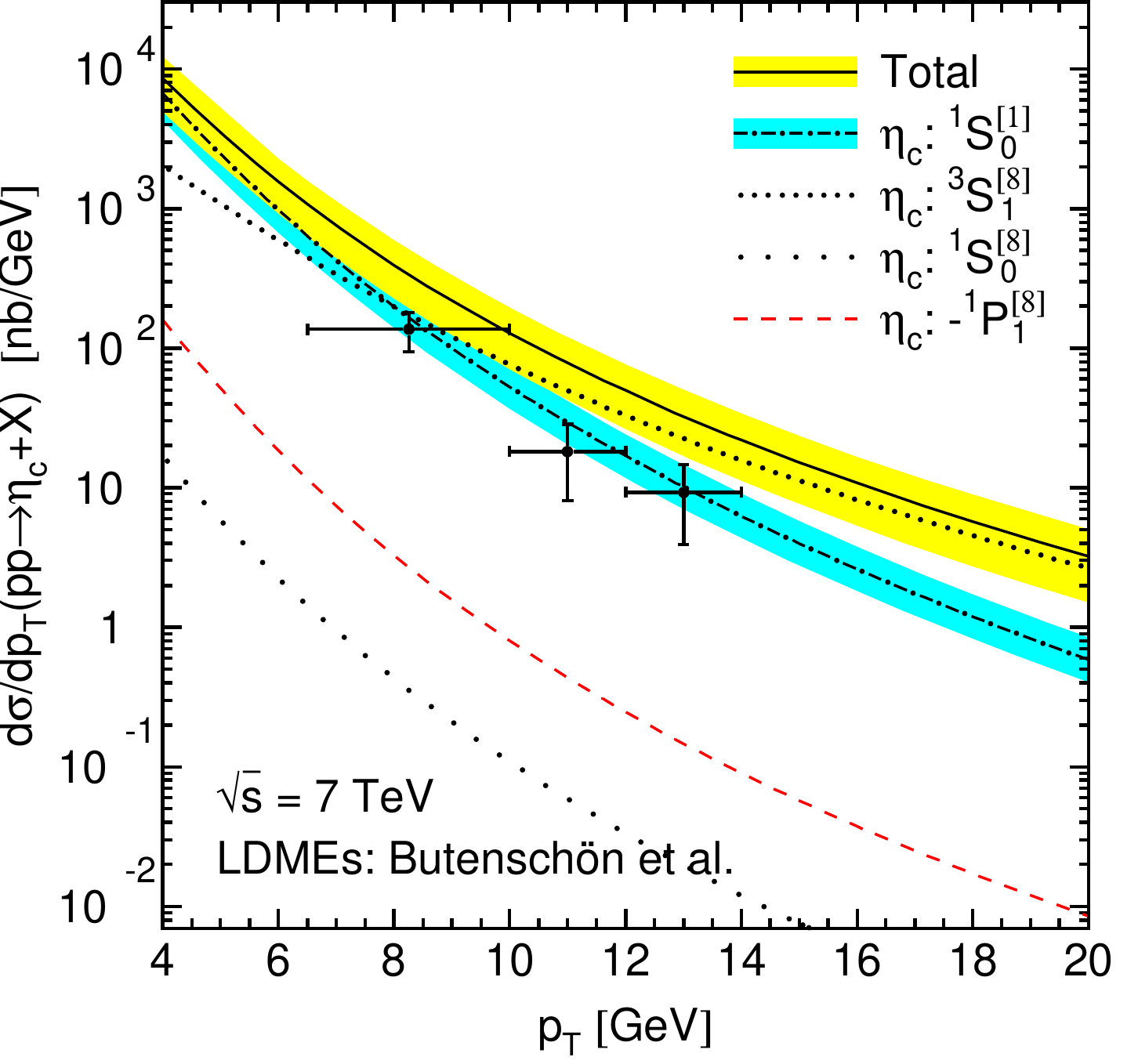}
  \hspace*{5mm}
 \includegraphics[width=0.3\textwidth]{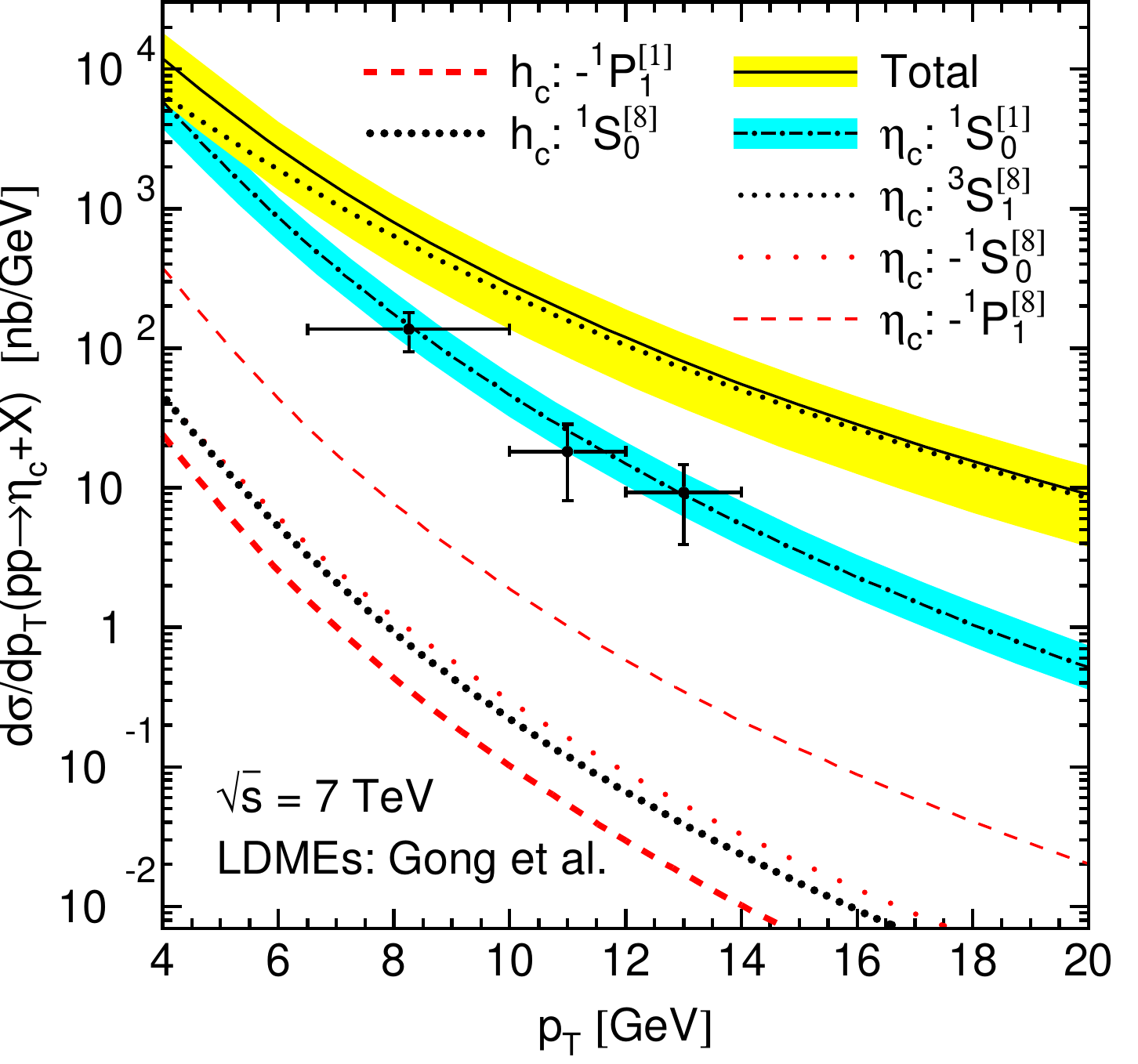}
 \end{center}
 \vspace*{-.5cm}
 \caption{ The LHCb \cite{LHCb:2014oii} measurements for prompt $\eta_c$ hadroproduction at $\sqrt{s}=7~\mathrm{TeV}$ are compared with the NLO NRQCD calculations evaluated with three different LDME sets from refs. \cite{Chao:2012iv,Butenschoen:2012px,Gong:2012ug}. Figures taken from
Ref. \cite{Butenschoen:2014dra}. \label{fig:eta}}
 \vspace*{0.cm}
\end{figure}

Due to the approximate heavy quark spin symmetry (HQSS) of NRQCD \cite{Bodwin:1994jh}, the production of $J/\psi$ is closely related to the production of $\eta_c$ by the following relations
\begin{align}
\langle \mo^{\eta_c}(\COcSa)\rangle &=\langle \mo^{J/\psi}(\COaSz)\rangle,\\
 \langle \mo^{\eta_c}(\COaSz)\rangle &=\frac{1}{3}\langle \mo^{J/\psi}(\COcSa)\rangle,\\
  \langle \mo^{\eta_c}(\COaPa)\rangle&=\frac{3}{2J+1}\langle \mo^{J/\psi}(\COcPj)\rangle.
\end{align}
Then the measurement of $\eta_c$ can provide a further test of the $J/\psi$ LDMEs.
The first prompt $\eta_c$ hadroproduction cross section was measured in 2014 by the LHCb collaboration with the centre-of-mass energies at $\sqrt{s}=7$~TeV and $8$~TeV \cite{LHCb:2014oii}.
With the three sets of the  $J/\psi$ LDMEs above,
the authors of ref. \cite{Butenschoen:2014dra} found that theoretical calculations overshoot experimental measurement, as shown in Fig.~\ref{fig:eta}.
Detailed analysis shows the LHCb data are almost saturated by the contribution from the CS $\CSaSz$ channel.
 This gives a constraint on $\langle \mo^{J/\psi}(\COaSz)\rangle$. By assuming the
data are completely contributed from the $\COcSa$ channel, the authors of ref. \cite{Han:2014jya} obtain an upper bound for $\langle \mo^{\eta_c}(\COcSa)\rangle$
\begin{align}
\langle \mo^{\eta_c}(\COcSa)\rangle < 1.46\times 10^{-2} \mathrm{GeV}^3.
\end{align}
This bound is consistent
with a previous study by the same authors on the $J/\psi$ yield and polarization \cite{Shao:2014yta}, so they argue that the prompt production of $\eta_c$ and
$J/\psi$ can be understood in the same theoretical framework.
The authors of ref.~\cite{Zhang:2014ybe} fit both the CO and CS LDMEs and their predictions are also compatible with data.
Recently, the LHCb collaboration reports their measurement of the $\eta_c$ hadroproduction cross section at $\sqrt{s}=13$~TeV \cite{LHCb:2019zaj}, which is consistent with previous data.

Within the same framework as that for the $J/\psi$ production, the study of $\Upsilon$ production at NLO was carried out by two groups \cite{Gong:2013qka,Feng:2015wka,Han:2014kxa}. Their results
show a slightly transverse polarization at large $p_T$, consistent with the measurements by CMS \cite{CMS:2012bpf} within
experimental uncertainties.

\begin{figure}[htb!]
 \begin{center}
 \vspace*{0.8cm}
 \hspace*{-5mm}
 \includegraphics[width=0.6\textwidth]{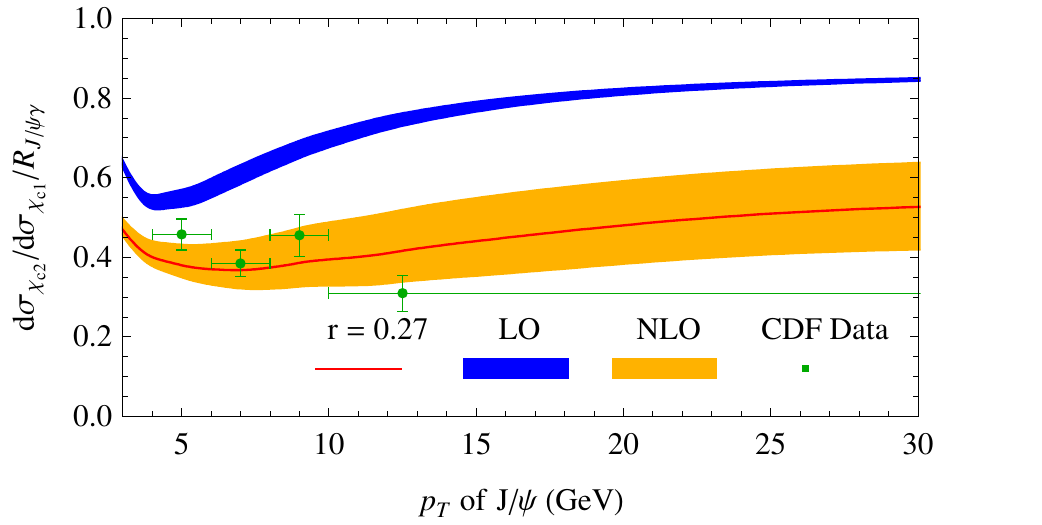}
 \end{center}
 \vspace*{-.5cm}
 \caption{ Transverse momentum distribution of ratio
$R_{\chi_c}$ at the Tevatron with cut $|y_{\chi_{cJ}}|<1$. Figure taken from
Ref. \cite{Ma:2010vd}. \label{fig:chi}}
 \vspace*{0.cm}
\end{figure}

For P-wave quarkonia, the first complete NLO study of $\chi_{cJ}$ production was performed in ref. \cite{Ma:2010vd} in 2010. At NLO, the CS $\CScPj$ channels scale as $p_T^{-4}$, which give large contributions at high $p_T$. They also find
$\CScPa$ decreases slower than
$\CScPb$, so the measured ratio of $R_{\chi_c}$ at the Tevatron \cite{CDF:2007mqb} can be naturally explained (see Fig. \ref{fig:chi}). In 2016, the authors of ref. \cite{Zhang:2014coi} performed an
global analysis of the existing data  on $\chi_c$ hadroproduction from the Tevatron \cite{CDF:2007mqb} and the LHC \cite{LHCb:2012af,CMS:2012qwg,LHCb:2013ofo,ATLAS:2014ala}.  In the meantime, the polarization of the $\chi_c$ was also predicted \cite{Shao:2014fca,Faccioli:2018uik}, but the experimental measurement was not available until 2019 by the CMS collaboration \cite{CMS:2019jas}.
Current experimental data seem to be consistent with the NLO results.

As mentioned in the previous section, in the high $p_T$ region, the partonic differential cross sections in the NRQCD factorization formula in Eq. \eqref{eq:NRQCD-pp} contain large logarithms like $\textrm{ln}(p_T^2/m_Q^2)$, which could ruin the convergence of perturbative expansion. Thus resummations of these large logarithmic terms are necessary. This can be done by using FF+NRQCD factorization approach. Applying the FF+NRQCD factorization approach with LP approximation was carried out in refs. \cite{Bodwin:2014gia,Bodwin:2015iua}for the hadroproduction of $J/\psi$, $\chi_{cJ}$ and $\psi(2S)$ at high $p_T$ and good agreement with the measurements is obtained. It was also found that contributions from $\COcSa$ and $\COcPj$ channels should almost cancel with each other so the produced $J/\psi$ is almost unpolarized, which confirms the conclusion in refs. \cite{Chao:2012iv,Shao:2014yta}. This implies that the
qualitative results in the NLO NRQCD calculations are not changed by LP resummation.
It is an interesting question whether the NLP resummation could change this conclusion.

\subsection{Low $p_T$ heavy quarkonium production}

\begin{figure}[htb!]
 \begin{center}
 \vspace*{0.8cm}
 \hspace*{-5mm}
 \includegraphics[width=0.6\textwidth]{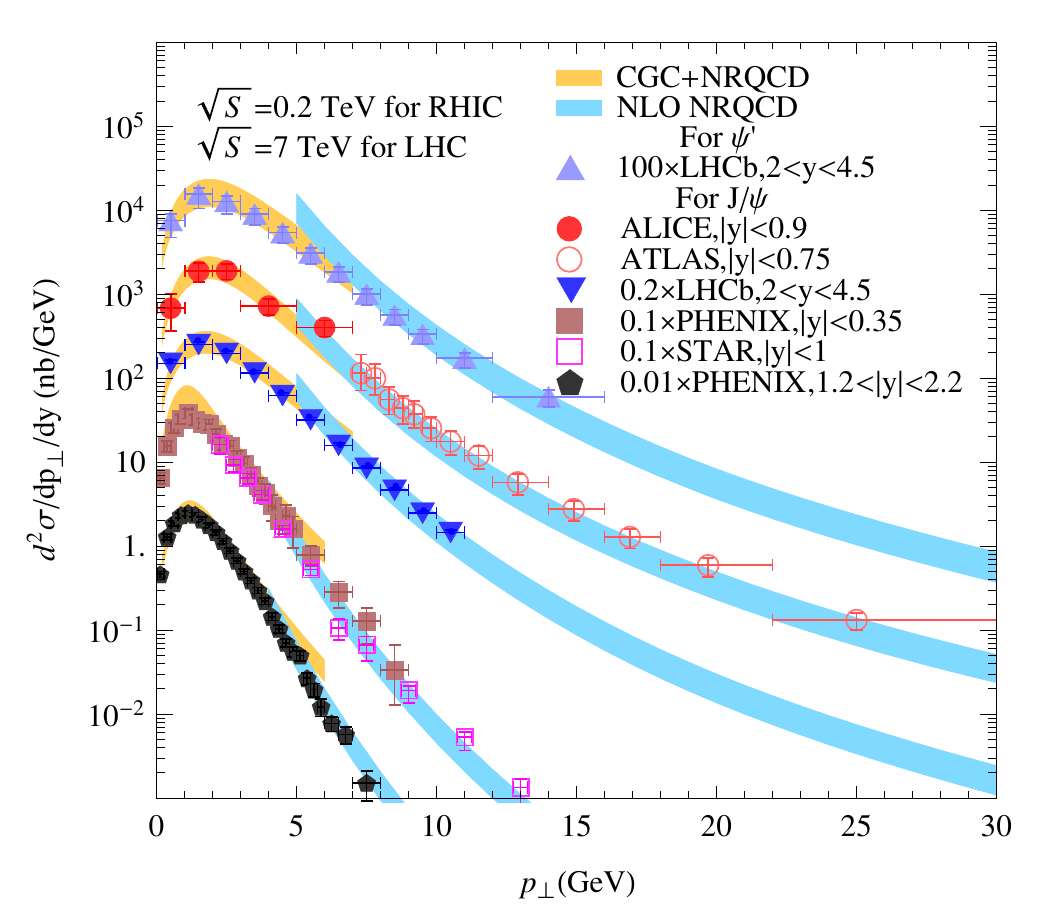}
 \end{center}
 \vspace*{-.5cm}
 \caption{The $\psi(2S)$ (top curve) and $J/\psi$ (other four curves) differential cross section as functions of $p_T$. Data from \cite{ATLAS:2011aqv,LHCb:2011zfl,PHENIX:2011gyb,ALICE:2011zqe,STAR:2012wnc,LHCb:2012geo}. Figure taken from
Ref. \cite{Ma:2014mri}. \label{fig:yield}}
 \vspace*{0.cm}
\end{figure}

\begin{figure}[htb!]
 \begin{center}
 \vspace*{0.8cm}
 \hspace*{-5mm}
 \includegraphics[width=0.6\textwidth]{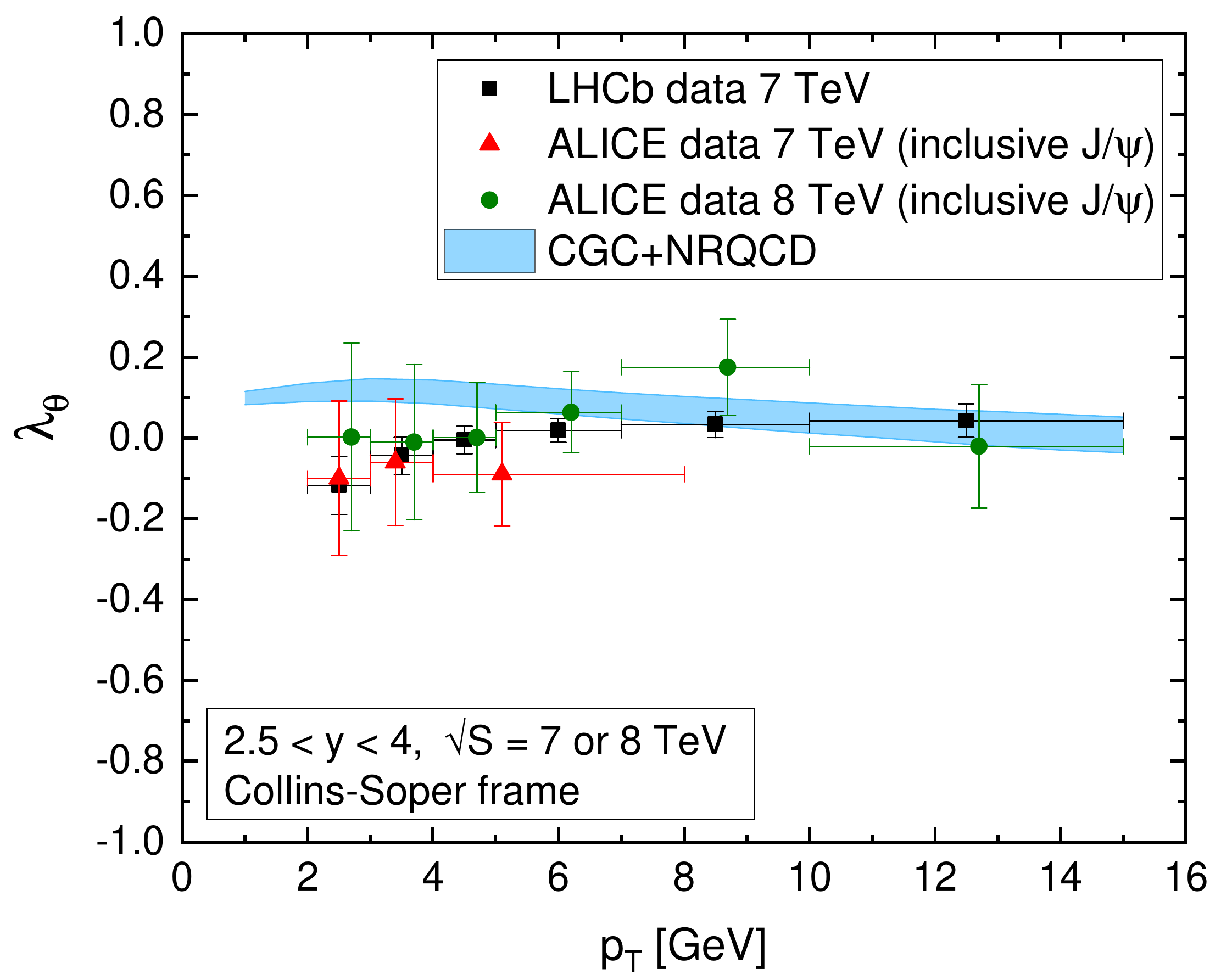}
 \end{center}
 \vspace*{-.5cm}
 \caption{The angular distribution coeffcients $\lambda_{\theta}$ in the Collins-Soper frame as functions of $p_T$. Data from \cite{ALICE:2011gej,ALICE:2018crw,LHCb:2013izl}. Figure taken from
Ref. \cite{Ma:2018qvc}. \label{fig:polar}}
 \vspace*{0.cm}
\end{figure}

At low $p_T \lesssim M$ ($M$ is the quarkonium mass) region, the collinear factorization formalism in Eq.~\eqref{eq:NRQCD-pp} is no longer applicable.
At this regime, large $\alpha_s \textrm{ln}(1/x)$ (small $x$, $x\sim M/\sqrt{s}$, where $\sqrt{s}$ is the collider center of mass energy) contribution arises at higher orders that may not be fully accounted for in collinear factorization framework.
Another source of $\mo(1)$ contribution is from the higher-twist multiparton matrix elements that are large at low $p_T$.
Both of these contributions can be computed systematically in the Color Glass Condensate (CGC) effective field theory~\cite{Iancu:2003xm,Gelis:2010nm}.
Combining the CGC and NRQCD formalisms, a new factorization framework for quarkonium production was proposed~\cite{Ma:2014mri,Kang:2013hta}, in which the SDCs in Eq.~\eqref{eq:NRQCD} are given by
\begin{align}\label{eq:dsktCO}
\begin{split}
\frac{d \hat{\sigma}_n}{d^2\vp d
y}\overset{\text{CO}}=&\frac{\alpha_s (\pi R_p^2)}{(2\pi)^{7}
(N_c^2-1)}
\int d^2 \vka  d^2\vk
\frac{\varphi_{p,y_p}(\vka)}{k_{1\perp}^2} \mathcal{N}_Y(\vk)\mathcal{N}_Y(\vp-\vka-\vk)
\,\Gamma^\kappa_8,
\end{split}
\end{align}
for the color octet channels and
\begin{align}\label{eq:dsktCS}
\begin{split}
\frac{d \hat{\sigma}_n}{d^2\vp d
y}\overset{\text{CS}}=&\frac{\alpha_s (\pi R_p^2)}{(2\pi)^{9}
(N_c^2-1)}
\int d^2\vka d^2\vk d^2\vkp
\frac{\varphi_{p,y_p}(\vka)}{k_{1\perp}^2}\\
&\hspace{-1.5cm}\times \mathcal{N}_{Y}(\vk)\mathcal{N}_{Y}(\vkp)\mathcal{N}_{Y}(\vp-\vka-\vk-\vkp)\,
{\cal G}^\kappa_1,
\end{split}
\end{align}
for the color singlet channels. In these expressions,  $\vp$ ($y$) is the transverse momentum (rapidity) of the produced heavy quarkonium, $y_p\equiv\ln(1/x_p)$ ($Y\equiv\ln(1/x_A)$) is the rapidity of gluons coming from dilute proton (dense proton), $\mathcal{N}_{Y}$ denotes the fundamental dipole amplitude, $\Gamma^\kappa_8$ and ${\cal G}^\kappa_1$ are the hard parts, $\varphi$ is an unintegrated gluon distribution inside the proton, and $\pi R_p^2$
is the effective transverse area of the proton. Such a CGC+NRQCD framework provides a good description of  $\psi(nS)$ yield~\cite{Ma:2014mri} and polarization at low $p_T$ \cite{Ma:2018qvc} in $pp$ collisions, shown in Fig. \ref{fig:yield} and Fig. \ref{fig:polar}.
Interestingly, the CGC+NRQCD result at small $p_T$ merges smoothly with the NLO NRQCD result at intermediate and large $p_T$, providing an unified description for quarkonium production in the full $p_T$ region. Note that, this framework is even more useful for quarkonium production in proton-nucleus collisions~\cite{Ma:2015sia,Ma:2017rsu,Ma:2017rsu,Ma:2018bax,Stebel:2021bbn}.

\section{Quarkonium production in $e^+e^-$ annihilation at B factories}

The quarkonium production in $e^+e^-$ annihilation at B factories is another important process  to test the NRQCD factorization.
For exclusive double charmonium production such as $e^+e^-\to J/\psi + \eta_c$, the first theoretical calculation at LO in both $\alpha_s$ and $v$ \cite{Braaten:2002fi,Liu:2002wq,Hagiwara:2003cw} gives a production cross section about $2.3\sim5.5$~fb.
It is much smaller than the measured value, which is $\sigma[J/\psi+\eta_c]\times B^{\eta_c}[\geq2]=(25.6\pm2.8\pm3.4)$~fb by Belle \cite{Belle:2002tfa}
and $\sigma[J/\psi+\eta_c]\times B^{\eta_c}[\geq2]=(17.6\pm2.8^{+1.5}_{-2.1})$~fb
by BaBar \cite{BaBar:2005nic}, where $B^{\eta_c}[\geq2]$ is the branching fraction for the
$\eta_c$ decaying into at least
two charged tracks.
Later, the authors of refs. \cite{Zhang:2005cha,Gong:2007db} show that the NLO QCD correction can substantially enhance
the cross section with a $K$ factor (the ratio of NLO to
LO ) of about $1.8\sim2.1$.
Meanwhile, the relative $O(v^2)$ correction is also found to be significant \cite{Braaten:2002fi,He:2007te,Bodwin:2007ga}.
Including both the $\alpha_s$ and $v^2$ corrections may resolve the large
discrepancy between theory and experiment.
Recently, the NNLO QCD correction to $e^+e^-\to J/\psi + \eta_c$ has been completed \cite{Feng:2019zmt}, which gives a
state-of-the-art calculation consistent with the BaBar measurement.

For inclusive $J/\psi$ production, the first measurements of the cross section are released
by the BaBar \cite{BaBar:2001lfi} and Belle \cite{Belle:2001lqi,Belle:2002tfa} collaborations. It was found by Belle \cite{Belle:2002tfa} that the cross section $\sigma[e^+e^-\to J/\psi + c\bar c ]=(0.87^{+0.21}_{-0.19}\pm0.17)$~fb
is about a factor of $5$ larger than the LO NRQCD
factorization predictions
including both the CS \cite{Cho:1996cg,Yuan:1996ep,Baek:1998yf,Kiselev:1994pu,Liu:2003jj} and CO \cite{Liu:2003jj} contributions.
Later, the Belle collaboration reported an updated
measurement: $\sigma[e^+e^-\to J/\psi + c\bar c ]=(0.74\pm0.08^{+0.09}_{+0.08})\mathrm{fb}$ \cite{Belle:2009bxr}, which is
smaller than the previous one.
But it is still much larger than
the LO NRQCD calculation.
The large gap between experiment and theoretical calculations is reduced by including the NLO QCD correction, which substantially enhances the cross section with a $K$ factor of about $1.8$~\cite{Zhang:2006ay}.

\begin{figure}[htb!]
 \begin{center}
 \vspace*{0.8cm}
 \hspace*{-5mm}
 \includegraphics[width=0.6\textwidth]{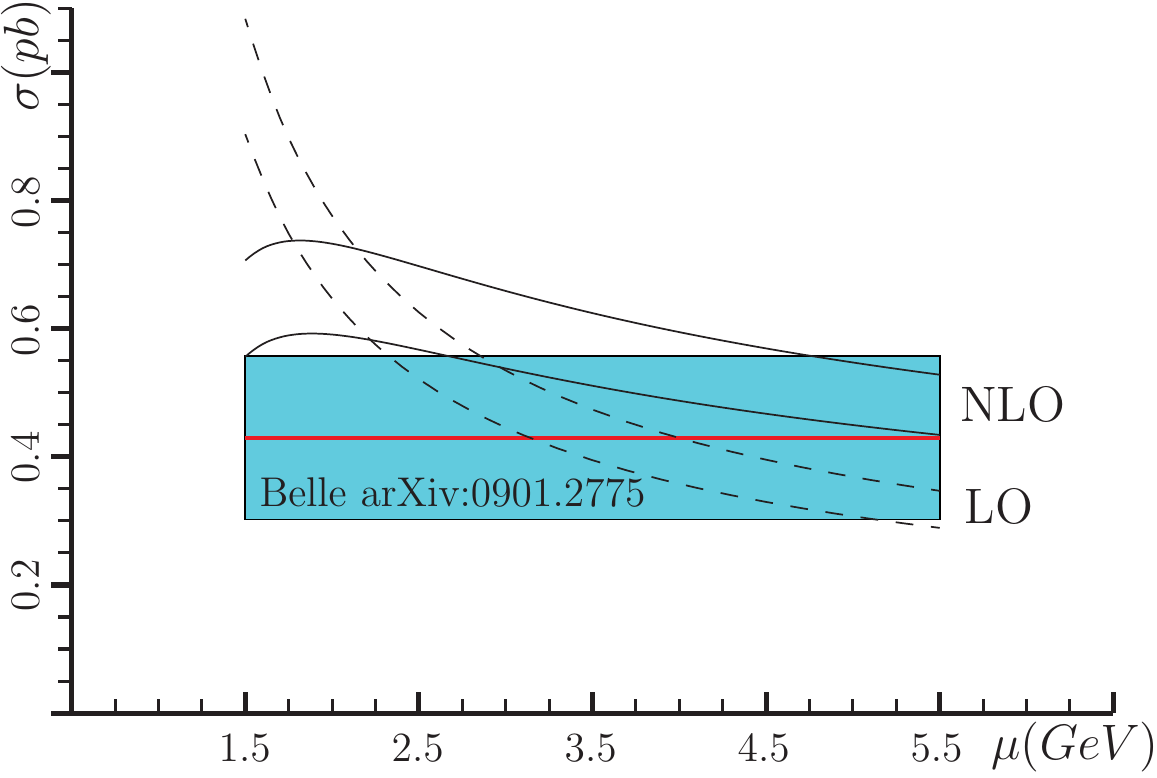}
 \end{center}
 \vspace*{-.5cm}
 \caption{Prompt cross sections of $e^+e^-\to J/\psi + gg $ as functions
of the renormalization scale $\mu$ at LO and NLO. The
upper curves correspond to $m = 1.4~\mathrm{GeV}$, and the lower ones
correspond to $m = 1.5~\mathrm{GeV}$. Figure taken from
Ref. \cite{Ma:2008gq}. \label{fig:ee}}
 \vspace*{0.cm}
\end{figure}

To explain $\sigma[e^+e^-\to J/\psi + X_{\mathrm{non}-c\bar c} ]$ measured by Belle \cite{Belle:2009bxr}, the NLO QCD correction to the CS channel $e^+e^-\to J/\psi + gg $ is calculated in refs. \cite{Ma:2008gq,Gong:2009kp}, which increases the LO
result by about $20\sim 30\%$.
As shown in Fig. \ref{fig:ee}, the NLO CS contribution saturates the Belle measurement.
The $O(v^2)$ relativistic correction \cite{He:2009uf,Jia:2009np} and the QED initial state radiation (ISR) effect \cite{Shao:2014rwa} have also been considered, which enhance the LO cross section by a factor of $20\sim 30\%$ and $15\sim 25\%$ respectively.
Surprisingly, including all these corrections leads to the CS contribution somewhat above the measurement by Belle, leaving little or no
room for the contribution of CO channel $e^+e^-\to J/\psi(\COcPj,\COaSz) + g$.
This provides an upper limit of the CO LDMEs.
By assuming a vanishing CS contribution and including the NLO QCD correction to $\sigma[e^+e^-\to J/\psi(\COcPj,\COaSz) + g]$, the authors of ref. \cite{Zhang:2009ym} obtain
\begin{align}
M_{0,4.0}^{J/\psi}<(2.0\pm0.6)\times 10^{-2}\text{GeV}^3,
\end{align}
which is much smaller than the value of CO matrix element  extracted from  hadron colliders, i.e. $M_{0,4.0}^{J/\psi} \approx 7.4\times 10^{-2}~\mathrm{GeV}^3$ \cite{Ma:2010yw,Ma:2010jj,Shao:2014yta}, bring the universality of LDMEs into question.

The universality problem and the polarization puzzle discussed in the previous section are two outstanding problems in NRQCD factorization. A possible solution is to resum high order relativistic corrections, since for charmonium $v^2\sim 0.3$ is not a small number. This is exactly the motivation of the SGF approach. A detailed and comprehensive phenomenological study for inclusive quarkonium production in SGF framework could help to understand these problems.
In the meantime, more experimental data with smaller uncertainties are also indispensable.

\section{Quarkonium production in $ep$ and $\gamma\gamma$ collisions}

In the photoproduction of charmonia in $ep$ collisions at
HERA, a quasi-real photon $\gamma$
emitted
from the incoming electron $e$ interacts with
a parton $i$ from the proton $p$ and produces a $c\bar c$ pair that evolves
into a charmonium state. There are two types of processes contributing to the photoproduction cross sections.
The first is the direct photo-production, in which the virtual photon interacts with the parton $i$ electromagnetically. The second is the resolved photo-production, in which the virtual photon
emits a parton, which then interacts with the parton $i$.
Combining
collinear factorization and NRQCD factorization, the inclusive $J/\psi$ photoproduction cross section can be written in the form of \cite{Butenschoen:2009zy,Butenschoen:2011yh},
\begin{align}\label{eq:NRQCD-ep}
d\sigma_{ep\to J/\psi+X}
 &=\sum_{i,j,n}\int dx_1dx_2dy  f_{\gamma/e}(x_1)f_{j/\gamma}(x_2)f_{i/A}(y) d\hat\sigma_{ij\to c\bar c[n]+X} \langle \mo^{J/\psi}_n\rangle,
\end{align}
where $f_{\gamma/e}(x_1)$ is the photon flux function, $f_{j/\gamma}(x_2)$ is either $\delta_{j\gamma}\delta(1-x_2)$ or the
PDF of parton $j$ in the resolved photon, $f_{i/A}(y)$ is
the PDF of the proton, and $d\hat\sigma_{ ij\to c\bar c[n]+X}$ is
the partonic cross section.

\begin{figure}[htb!]
 \begin{center}
 \vspace*{0.8cm}
 \hspace*{-5mm}
 \includegraphics[width=0.45\textwidth]{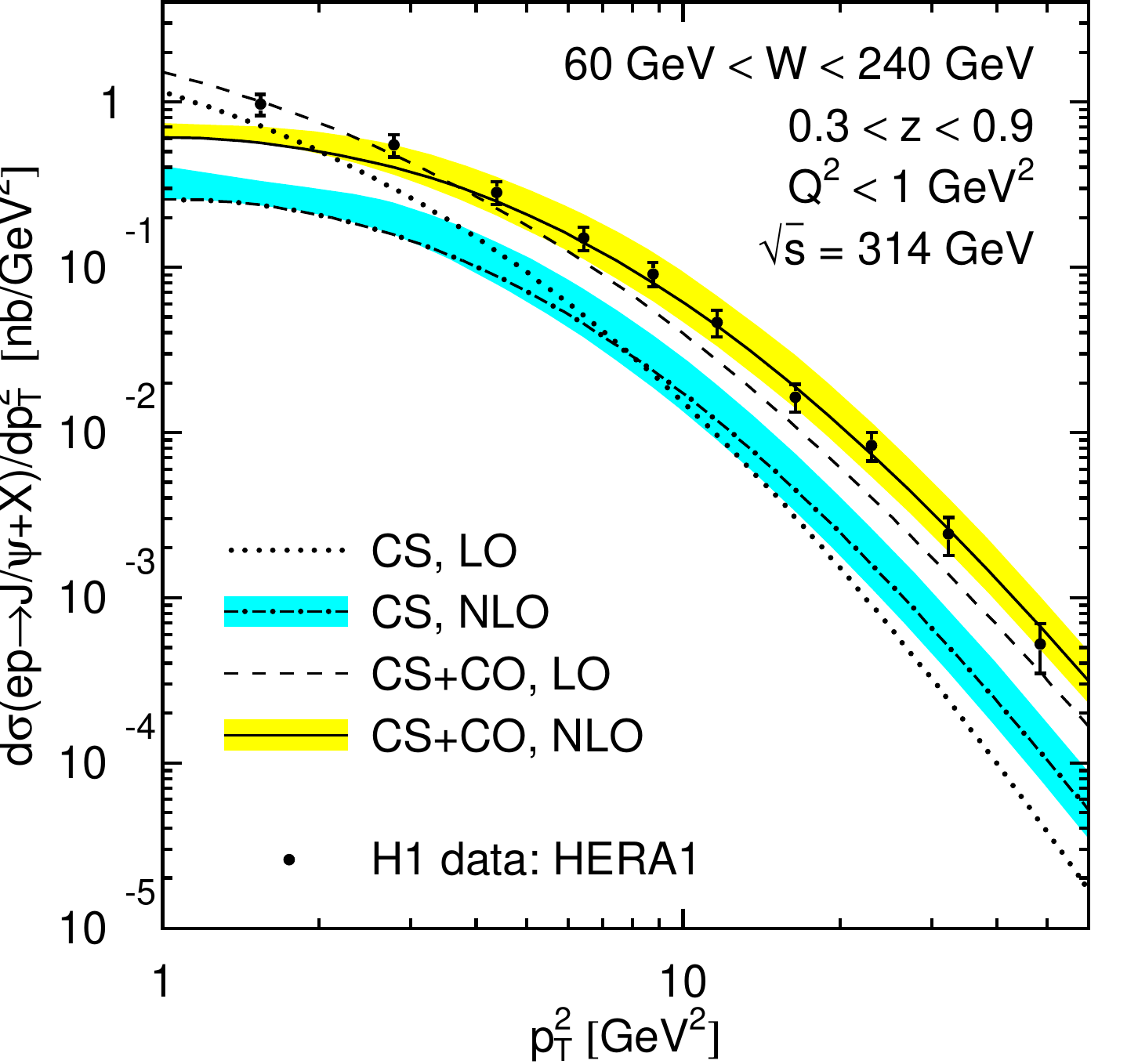}
 \hspace*{5mm}
 \includegraphics[width=0.45\textwidth]{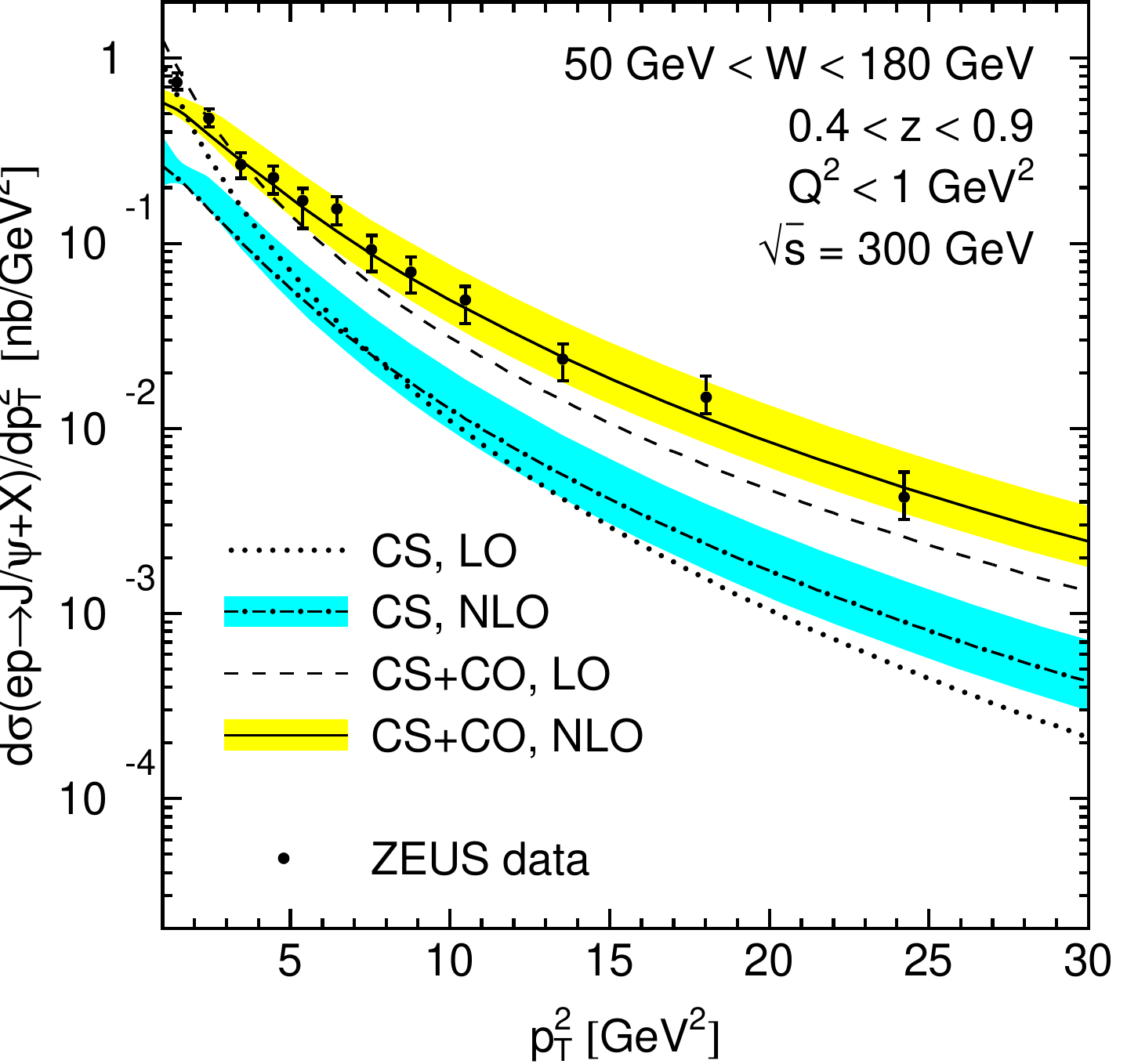}
 \end{center}
 \vspace*{-.5cm}
 \caption{NLO NRQCD calculation compared to HERA data. Figures taken from
Ref. \cite{Butenschoen:2011yh}. \label{fig:jpsi2}}
 \vspace*{0.cm}
\end{figure}

The NLO QCD correction to the partonic cross sections $d\hat\sigma_{ i\gamma\to c\bar c[n]+X}$ for the
direct production was first performed by Kr\"{a}mer in 1995 \cite{Kramer:1995nb}. In 2009, complete NLO calculation was obtained in refs. \cite{Butenschoen:2009zy,Artoisenet:2009xh,Chang:2009uj}. As shown in Fig.~\ref{fig:jpsi2}, the yield data from HERA are roughly consistent with the global fit at NLO performed in ref. \cite{Butenschoen:2010rq,Butenschoen:2011yh}.

\begin{figure}[htb!]
 \begin{center}
 \vspace*{0.8cm}
 \hspace*{-5mm}
 \includegraphics[width=0.45\textwidth]{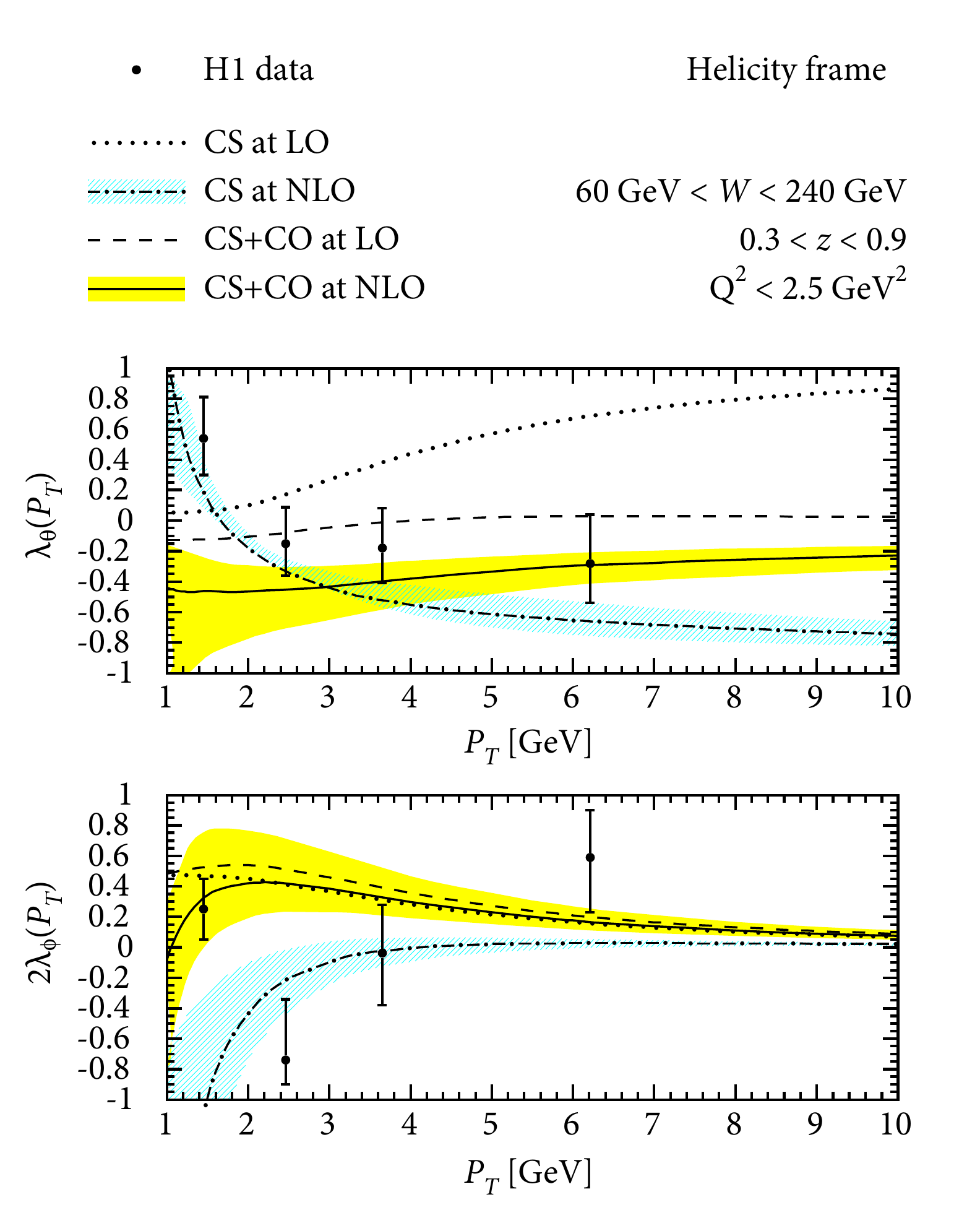}
 \hspace*{5mm}
 \includegraphics[width=0.45\textwidth]{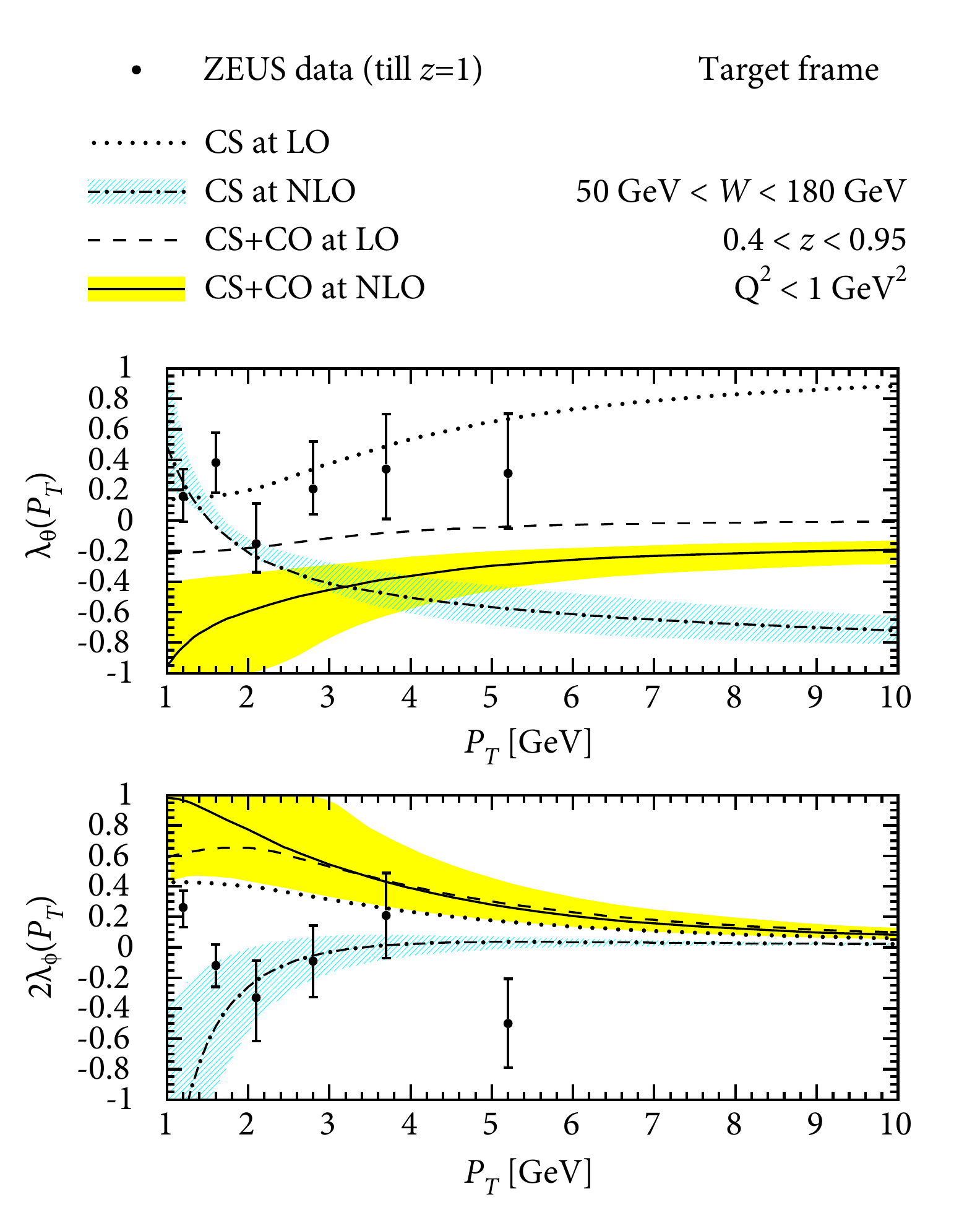}
 \end{center}
 \vspace*{-.5cm}
 \caption{NLO NRQCD calculations for polarization observables as functions
of $p_T$ compared to $\mathrm{H1}$ \cite{H1:2010udv} and ZEUS \cite{ZEUS:2009qug} data. Figures taken from
Ref. \cite{Lansberg:2019adr}. \label{fig:jpsiphotopol}}
 \vspace*{0.cm}
\end{figure}

In addition to the unpolarized $J/\psi$ yield, the polarization of $J/\psi$ photoproduction has also been studied \cite{Artoisenet:2009xh,Chang:2009uj,Butenschoen:2011ks}.
Fig. \ref{fig:jpsiphotopol} shows
the LO and NLO NRQCD results of polarization observables as a function of $p_T$ by ref. \cite{Butenschoen:2011ks} comparing to two HERA datasets. It is found that NRQCD predicts the $J/\psi$ produced at large $p_T$
to be approximately unpolarized, both at LO and NLO, which is confirmed by the $\mathrm{H1}$ data (left panel). However, the ZEUS measurement (right panel) exhibits a tendency towards transverse polarization.
A possible explanation is that diffractively produced
vector mesons prefer to be strongly transversely polarized in the endpoint region $z\approx 1$ \cite{Butenschoen:2011ks}.

\begin{figure}[htb!]
 \begin{center}
 \vspace*{0.8cm}
 \hspace*{-5mm}
 \includegraphics[width=0.5\textwidth]{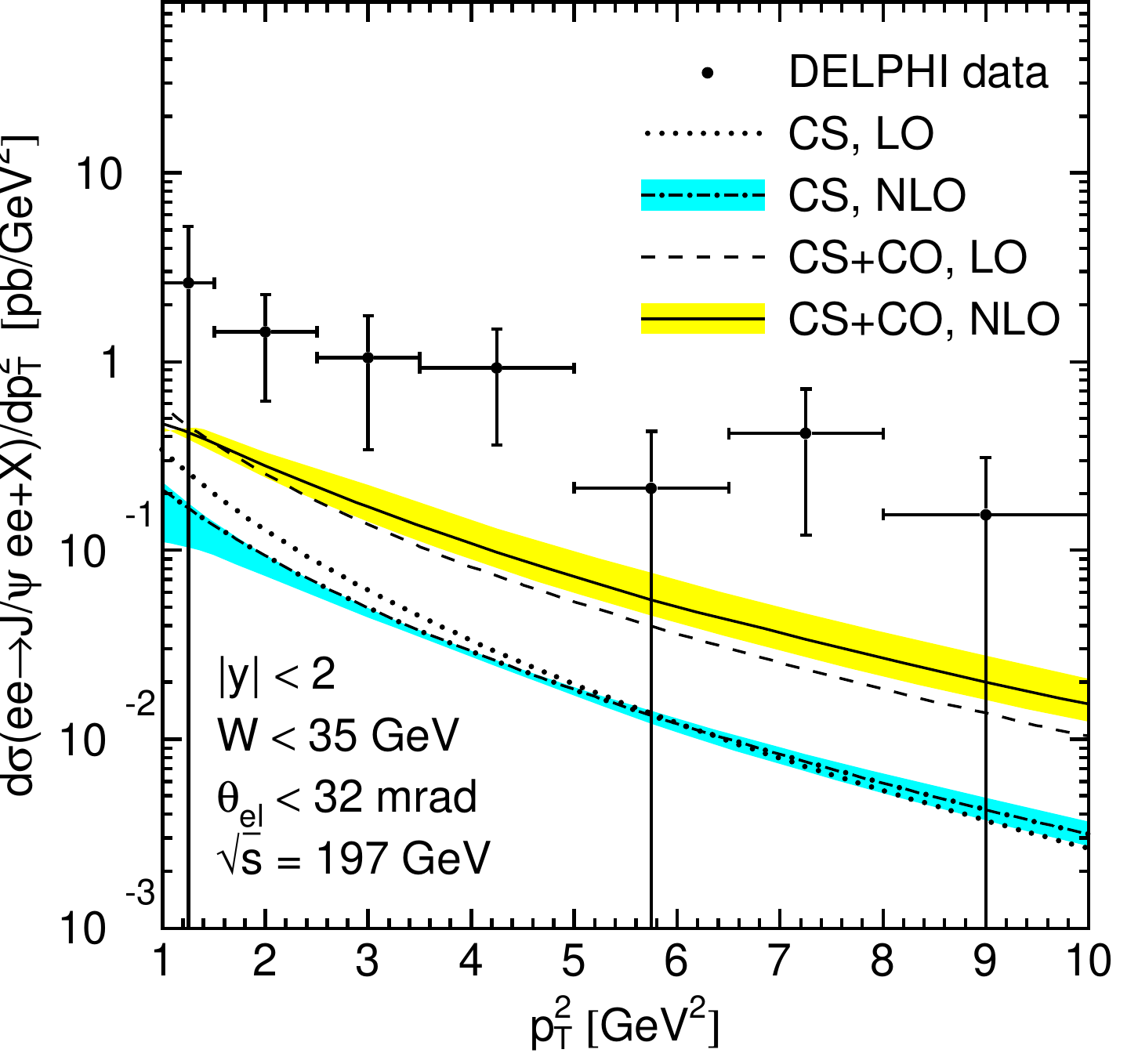}
 \end{center}
 \vspace*{-.5cm}
 \caption{NLO NRQCD calculation compared to DELPHI data. Figure taken from
Ref. \cite{Butenschoen:2011yh}. \label{fig:rr}}
 \vspace*{0.cm}
\end{figure}

The inclusive cross section for $J/\psi$ production in $e^+e^-$ colliders via $\gamma \gamma$ fusion has also been measured by the DELPHI collaboration \cite{DELPHI:2003hen}. Similar to the photoproduction, both direct photon and resolved photon contribute to the cross section. The first complete NLO NRQCD computation including the resolved contributions was studied in ref. \cite{Butenschoen:2011yh}. However, by using the LDMEs obtained from the global fit, the predicted cross sections is several times below the DELPHI data, as shown in Fig. \ref{fig:rr}.
This is caused by a cancellation between the $\COaSz$ and $\COcPj$ contributions owing to the negative value of
$\langle \mo^{J/\psi}(\COcPz)\rangle$ obtained by the global fit \cite{Lansberg:2019adr}. Thus a global analysis of the world $J/\psi$ data is still challenging.

\section{Summary}

In this article, we have reviewed some theoretical methods to describe heavy quarkonium production, including the color evaporation model, color singlet model, NRQCD factorization, fragmentation function approach, and soft gluon factorization. We then emphasize the current status of the phenomenological study of charmonium production, mainly in the NRQCD factorization framework. We concentrate on the comparison between theoretical predictions and experimental
measurements for the charmonium production in four important processes: $pp$ collision, $e^+e^-$ annihilation, $ep$ collision and $\gamma\gamma$ collision. After the NLO contributions are taken
into account, the NRQCD factorization can give a
qualitatively correct description for quarkonium production.
 Especially, by combining NRQCD factorization with CGC effective theory, an unified description for quarkonium hadroproduction in full $p_T$ region is obtained. However, LDMEs determined from different choices of datasets can disagree with one another, and none of them are able to give a global description of all important observables, such as the total yield, momentum differential yield and polarization. The polarization puzzle and the universality problem are two outstanding problems in NRQCD factorization, which may be caused by the bad convergence of velocity expansion. Hopefully, these difficulties could be resolved or relieved in the SGF framework with well controlled relativistic corrections.

\begin{acknowledgments}
The work is supported by the National Natural Science Foundation of China (Grants No. 11875071, No. 11975029), the National Key Research and Development Program of China under Contracts No. 2020YFA0406400, and the Qilu Youth Scholar Funding of Shandong
University.
\end{acknowledgments}

\providecommand{\href}[2]{#2}\begingroup\raggedright\endgroup



\end{document}